\documentclass[twocolumn]{emulateapj}
\usepackage[caption=false]{subfig}

\newcommand{\Rsun}{R$_\odot$}
\newcommand{\Msun}{M$_\odot$}
\newcommand{\Mdotu}{\Msun$\,$yr$^{-1}$}
\def\lsim{\mathrel{\rlap{\lower4pt\hbox{\hskip1pt$\sim$}}
    \raise1pt\hbox{$<$}}}                % less than or approx. symbol
\def\gsim{\mathrel{\rlap{\lower4pt\hbox{\hskip1pt$\sim$}}
    \raise1pt\hbox{$>$}}}                % greater than or approx. symbol

\slugcomment{Manuscript \today}

\shorttitle{Understanding the Unusual X-Ray Emission of the Massive Close Binary WR~20a}
\shortauthors{Montes et al.}

\begin{document}

\title{UNDERSTANDING THE  UNUSUAL X-RAY EMISSION PROPERTIES OF THE MASSIVE, CLOSE BINARY WR 20A: A HIGH ENERGY WINDOW INTO THE STELLAR WIND INITIATION REGION}

\author{ Gabriela Montes\altaffilmark{1,2}, Enrico Ramirez-Ruiz\altaffilmark{1}, Fabio De Colle\altaffilmark{1,3} and Rachel Strickler\altaffilmark{1}}
\altaffiltext{1}{Department of Astronomy and
  Astrophysics, University of California, Santa Cruz, CA
  95064}
\email{gmontes@ucsc.edu}
\altaffiltext{2}{Instituto de Astrof\'isica de Andaluc\'ia (IAA), CSIC, Glorieta de la Astronom\'ia s/n, E-18008, Granada }
               
\altaffiltext{3}{Instituto de Ciencias Nucleares, Universidad Nacional Aut{\'o}noma de M{\'e}xico, A. P. 70-543 04510 D. F. Mexico}

\begin{abstract}
The problem of explaining the X-ray emission properties of the massive, close binary  WR~20a is discussed. 
Located near the cluster core of Westerlund 2, WR~20a is composed of two nearly identical Wolf-
Rayet stars of 82 and 83 solar masses orbiting with a period of only 3.7 days.
Although {\it Chandra} observations were taken during the secondary optical eclipse, the X-ray light curve shows no signs of a  flux decrement. In fact, WR 20a appears slightly more X-ray luminous and softer during the optical eclipse, opposite to what has been  observed in other  binary systems.   To aid in our interpretation of the data, we compare with the results  of  hydrodynamical simulations using the adaptive mesh refinement code {\it Mezcal} that includes radiative cooling and a radiative acceleration force term. 
It is shown that the X-ray emission can be successfully  explained in models where the wind-wind collision interface in this system occurs while the outflowing  material is still being accelerated. 
Consequently, WR 20a serves as a critical test-case for how radiatively-driven stellar winds initiate and interact.  Our models  not only procure a robust description of  current  {\it Chandra} data, which cover the orbital phases between 0.3 to 0.6, but  provide detailed predictions  over the entire orbit.
\end{abstract}


\section{Introduction} 
Among the most massive stars, which tend also to be the hottest and most luminous, the emanating stellar winds can be very strong, with important consequences for both the star's own evolution as well as for the character of the emitting radiation. 
There is extensive evidence that hot-star winds are not the smooth, steady outflows idealized in the simple CAK theory
\citep{castor75}
but rather have extensive substructure  \citep{lucy80,owocki2011}. 
Massive O-type and Wolf-Rayet (WR) stars commonly show soft X-ray emission  \citep{naze09, skinner12} and sometimes also non-thermal radio emission \citep{abbott86,cappa04,montes09}, both of which are thought to originate from embedded wind shocks.
In binary systems,   the kinetic energy of the winds can be efficiently reconverted into radiation by  shocks which form when the  stellar winds collide \citep{usov92}.  The resultant  hard ($\gsim 2$~keV) X-ray component  \citep[e.g.,][]{stevens92} usually dominates the total X-ray luminosity, greatly exceeding the empirical relation log($L_x/L_{\rm bol}) \sim -6.9$ found for single hot massive stars \citep{chlebowski89, sana06}. 
The study of the X-ray emission from colliding binaries represents an indirect way of studying the properties of the stellar wind itself \citep[e.g.,][]{pittard02}. The hardness of the emission is related to the shock temperature which in turns depends on the pre-shock velocity of the wind. The X-ray brightness, on the other hand, is sensitive to the pre-shock density of the wind, and hence on the mass loss rate.

The theory for winds emanating from hot stars is mature enough that it is now finding application in many more complex circumstances. One example is colliding winds of close, massive star binaries. In several cases these involve  massive O-type and WR  members  with separations of only a few stellar radii. In these systems,  a wind-wind interface can occur   while material  is still being  accelerated 
\citep{antokhin04, pittard09}. 
This suggests that observations of very close massive binaries could help understand  how massive stars drive  winds. However, most of the X-ray studies  in massive binaries have been performed for systems with orbital periods of months to years 
\citep[e.g., WR~140,][]{pollock05}, and the relationship between the  X-ray emission and the wind properties in short-period systems has been difficult to determine because  of a variety of added complications  such as wind absorption and strong radiative interactions.

WR~20a is a close massive binary system well-suited to study how radiatively-driven stellar winds initiate and interact.
This eclipsing binary system is composed of  two almost identical stars (82+83~$M_\odot$; WN6ha+WN6ha)   in a circular orbital with a period of  $\sim 3.7$~days \citep{bonanos04, rauw05}.  X-ray emission has been detected for this system \citep{naze08} with a $L_x/L_{\rm bol}$ ratio greatly exceeding that observed for single WN stars \citep{skinner12}. What is more, the X-ray spectrum shows evidence of a hot component around 1.3-2.0~keV, supporting the idea  that the  X-ray emission arises from the wind-wind  interaction region.

An intriguing characteristic of the X-ray lightcurve resulting from WR~20a is the lack of observed eclipses at both soft and hard X-ray energies, which are, however,  evident in the optical lightcurve. 
Interestingly,  the X-ray emission shows an  increase in brightness, which is   slightly more pronounced  at softer energies (10-20\%), when the wind-wind interaction region is seen face-on. As a result, WR~20a appears to be brighter and somewhat softer during the secondary optical eclipse \citep{naze08}.  
These features  are very different from the X-ray lightcurves that have been observed and predicted in other  systems \citep[e.g.][]{pittard10},  where a clear X-ray luminosity decrement  is seen when the wind-wind interaction region  gets occulted  by the stars. 
One possible explanation for this unique behavior could be the orbital modulation of the absorbing wind column density, although its exact temporal dependence  will be determined  by the structure of the flow near the complex  wind-wind interaction region.

In this paper we present two-dimensional, axisymmetric 
hydrodynamical simulations of the wind-wind interaction region in WR~20a and test two possible scenarios: 
one in which  the emanating winds  are assumed to have achieved their terminal velocities
before interacting and another one in  which   the winds are still accelerating.
The results of the simulations are then  used to compute the X-ray emission, which are in turn  compared with   {\it Chandra} observations in order to 
 help  answer key remaining questions about the size and structure of the shocked and unshocked regions. Our accelerating wind model for the close interacting   binary WR~20a not only provides an accurate description of  current  X-ray observations, which cover the orbital phases between 0.3 to 0.6, but also give detailed predictions of the expected  lightcurve  over the entire orbital period.

\section{Modeling Wind-Wind Interactions in WR20A}
\label{sec:model}

\begin{table}
\begin{center}
\caption{Properties of WR20A as derived by  1: \citet{bonanos04}, 2: \citet{rauw05}, and 3: \citet{naze08} }
\label{table1}

\begin{tabular}{lcc}
\tableline 
\tableline \\
Parameter         & Value   & Reference\\
\tableline \\
Spectral Type      & WN6ha+WN6ha                        &  1        \\
$M$ (\Msun)        & 82.7 $\pm$5.5 + 81.9$\pm$5.5       &  1        \\
$R_\star$ (\Rsun)   & 19.3$\pm$0.5                       &  1        \\
$T_{\rm eff}$ (K)       & 43$\,$000$\pm$2000                 &  1        \\
$\dot{M}$ (\Mdotu) & 8.5$\times\,10^{-6}$                &  1         \\
$v_\infty$ (km$\,$s$^{-1}$) & 2800                        &  1        \\
$L_{\rm bol}$ ($L_\odot$) & 1.91$\times 10^6(d/8~{\rm kpc})^2$ & 3         \\
$L_{X}$ (erg$\,$s$^{-1}$)   & 5.17$\times 10^{33}(d/8~{\rm kpc})^2$ &  3         \\
\tableline\\
$P$ (days)                 & 3.686$\pm$0.01               &    2      \\
$a$ (\Rsun)                & 53.0$\pm$0.7                   &   2       \\
$i$                        & 74$^\circ$.5$\pm$2$^\circ$.0  &    2      \\
$d$ (kpc)                  &  2-8                         &   2       \\
\tableline
\end{tabular}
 \end{center}
\end{table}

\subsection{Hydrodynamical Models}

The modeling of stellar wind interactions in massive binaries is  challenging because  the geometry of the interacting region is highly dependent on  the ratio  of the  momentum of the emanating winds  and the eccentricity of the orbit \citep[e.g.][]{parkin09,okazaki08,corcoran11, corcoran10, parkin11}. 
The high degree of symmetry in WR~20a, which is composed of two almost identical stars in a near circular orbit,  allows for major simplifications to be made  in both  the hydrodynamical formalism and the  modeling of the  observed radiation (see Table~\ref{table1}).

To study the properties of the resultant X-ray lightcurves, we carried out two dimensional simulations of the wind interacting region using the hydrodynamics code {\it Mezcal} \citep{decolle06, decolle08, decolle12} including radiative cooling and a simplified radiative acceleration force term. 
We solve the following set of equations:
\begin{eqnarray}
  && \frac{\partial \rho}{\partial t} + \nabla \cdot (\rho \vec{v})=0 \;, \nonumber\\
  && \frac{\partial \rho \vec{v}}{\partial t} + \nabla \cdot (\rho \vec{v} \vec{v} + p \mathbb{I}) = f \;, \\
  && \frac{\partial e}{\partial t} + \nabla \cdot (e + p)\vec{v} = \rho \vec{f}\cdot\vec{v} - n^2 \Lambda (T) \;,
\label{eq:hydro}
\end{eqnarray}
where $\rho$ is the mass density, $\vec v$ is the velocity vector, 
$p$ is the thermal total pressure, $\mathbb{I}$ is the identity matrix, 
$e$ is the total energy defined as $e= \frac{1}{\gamma-1}p_{\rm gas}+ \frac{1}{2} \rho v^2$ 
(being $\gamma=5/3$ the adiabatic index), and $\Lambda(T)$ is the coronal ionization
equilibrium cooling function \citep{dalgarno72} adapted  to the abundance ratios of  WN6ha stars \citep{antokhin04}.

In the acceleration region, the wind density is given by 
\begin{equation}
\rho_{\rm w}(r) = {\dot{M} \over 4 \pi r^2 v(r)}, 
\end{equation}
while the  wind velocity law  is characterized here using a phenomenological $\beta$-law form: 
\begin{equation}
v(r) = v_0 + \left( v_\infty - v_0 \right) \left(1- {R_\ast \over r}\right)^\beta, 
\label{eq:beta}
\end{equation}
where $R_\ast$ is the stellar radius, 
$\beta$ is the velocity exponent, and $v_0$ is the velocity at the stellar 
surface, here taken for numerical convenience to be\footnote{We perform two additional simulations using $v_0=v_\infty/30$ and $v_0=v_\infty/100$ and found no appreciable changes within  the shocked wind region and  no discernible differences in the resulting X-ray lightcurves.}  $v_\infty/10$.

The density and velocity radial profiles described above are introduced in the simulation by adding a force 
term $f$ to the momentum equation:
\begin{equation}\label{eq:f}
 f = \frac{\dot{M} R_{\ast} \beta}{4 \pi r^4} \left( v_\infty - v_0 \right) \left( 1-\frac{R_\ast}{r} \right)^{\beta-1}
\end{equation}

In our simulations, we use a uniform grid of axial/radial size 
of $2 \times 10^{13}$~cm, with $640$ cells along both axis, 
corresponding to a resolution of $\approx 3 \times 10^{10}$~cm.
We impose reflecting boundary conditions at $z=0$ and  the wind is injected 
from a spherical boundary located at both stellar surfaces.

We run two models using the binary  and wind stellar  parameters listed  in Table~\ref{table1}. 
In the first one an instantaneous wind acceleration is imposed at the stellar boundary ({\it instant acceleration}) resulting in a constant wind velocity $v=v_\infty $ (i.e., setting $f=0$ in equation \ref{eq:f}). 
In the second one ({\it velocity profile}) we include a wind accelerating region with a $\beta$-law profile (equation \ref{eq:beta}) with $\beta=1$  \citep{rauw05},
resulting in a velocity $v (x=a/2) \approx 965$ km/s in the wind-wind interaction region\footnote{WR stars seems to follow a hybrid velocity structure with $\beta=1$ in the inner regions to a slower law with $\beta=5$ for the outer wind \citep{grafener05}.}.

\begin{figure}[]
%\centering
\includegraphics[height=1.4\linewidth,clip=true]{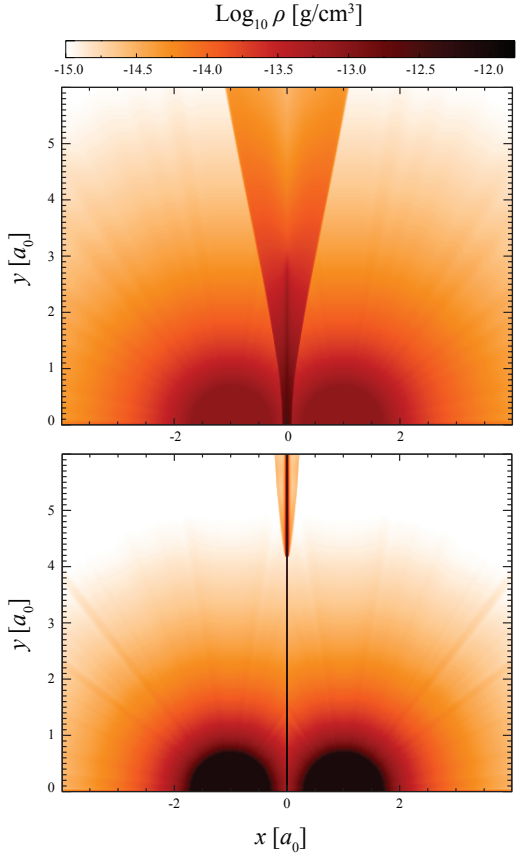}
\caption{Density structures resulting from the  instantaneous acceleration model ({\it top}), and from the  accelerating wind  model ({\it bottom}) with a $\beta$-law profile ($\beta=1$). $x$-$y$ axis are in units of $a_0=a/2$, which is the distance from one star to the contact discontinuity along the orbital plane. Following \citet{stevens92} we calculate the cooling parameter $\chi=t_{\rm cool}/t_{\rm dyn}$ as function of $y$, with $t_{\rm cool}=kT_s/4n_w \Lambda (T_s)$, where $n_w$ and $T_s$ are the pre-shock density and shock temperature respectively (both functions of $r$), and $t_{\rm dyn}=x_0/c_s$, with $x_0=a/2$ being the distance from one star to the contact discontinuity  and $c_s$ the sound speed. The cooling appears to be efficient ($\chi < 1$) up to $y/x_0 \approx 4$, which is reflected in the structure of the wind-wind interaction region shown in the {\it bottom} panel}.
\label{fig:hydro}
\end{figure}

\subsection{X-ray Lightcurves}

The results from the hydrodynamic simulations are used to construct both density and temperature profiles for the X-ray emitting material  along  the $y$-axis by averaging over  the width of the   shocked region ($x$-axis). The thermodynamical profiles are then used as initial conditions for the radiative transfer code {\tt Cloudy 13.00} \citep{ferland13} to determine emissivity and opacity profiles for the shocked wind gas in the 0.1 - 10~keV energy range with 240 logarithmically spaced energy bins. 
Assuming cylindrical symmetry we are able to construct three-dimensional  maps of the shocked and unshocked gas. 

The opacity for the unshocked  material  is calculated  using  {\tt Cloudy} under the assumption that the outflowing  wind from each star  is irradiated by a stellar blackbody chosen to match the observed properties of the binary (Table\ref{table1}). 
In addition to the stellar radiation field,  a central X-ray source  characterized by  a thermal bremsstrahlung  spectrum with $T_{\rm ff}$ and $L_{\rm ff}$ was also included. 
The values of  $T_{\rm ff}$ and  $L_{\rm ff}$ are chosen to closely match the total unabsorbed radiation  emitted by the shocked region, although as shown by \citet{antokhin04}, the opacity of the cold wind material  at the energy ranges where the X-ray lightcurves are constructed is not  sensitive to  the properties of the X-ray emission. 
Finally, X-ray lightcurves for the entire orbit  are constructed by ray-tracing the   emitted radiation  from the shocked region over the three-dimensional  wind gas maps, which are assumed to be inclined  with respect to the observer at an  angle $i$ (see Table~\ref{table1}).

\section{The Gas and X-ray Emission Properties}
\label{sec:results}

\subsection{The Wind-Wind Interaction  Region}

The results of our simulations are shown in Figure \ref{fig:hydro} for the two different models. 
In the {\it velocity profile} model, the  lower pre-shock velocities and the corresponding larger pre-shock densities 
result in a sizable increase in the post-shock densities (with respect to the {\it instant acceleration} model) which, together with the  accompanying lower post-shock temperatures lead to efficient radiative cooling 
in the inner regions. As shown in Figure \ref{fig:hydro}, this sizable increase in energy loss causes the shocked wind material to compress into a thin layer up to a height of about $y/x_0 \approx  4$ where  cooling becomes less efficient.

\begin{figure}[]
\centering
\includegraphics[height=1.2\linewidth,clip=true]{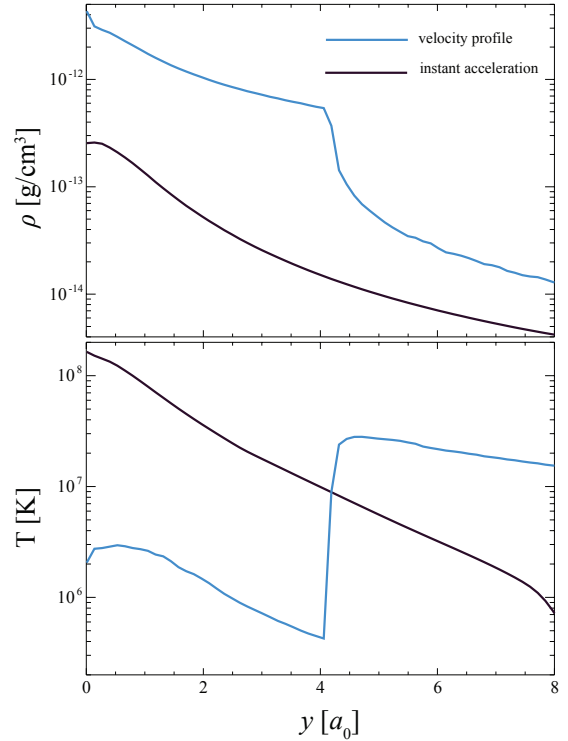}
\caption{Density and temperature profiles within the shocked material, obtained by averaging over the width of the  shocked region ($x$-axis), for the  {\it instant acceleration} (black line) and {\it velocity profile}  (blue line) models, respectively. At   $y/x_0 < 4$, a clear temperature inversion layer is produced  in the {\it velocity profile}  model as a result of the corresponding larger post-shock densities and  lower post-shock temperatures, which lead to more efficient radiative cooling  with respect  to the {\it instant acceleration} model.}
\label{fig:prof}
\end{figure}

\begin{figure*}[]
\centering\includegraphics[height=1.1\linewidth,angle=90,clip=true]{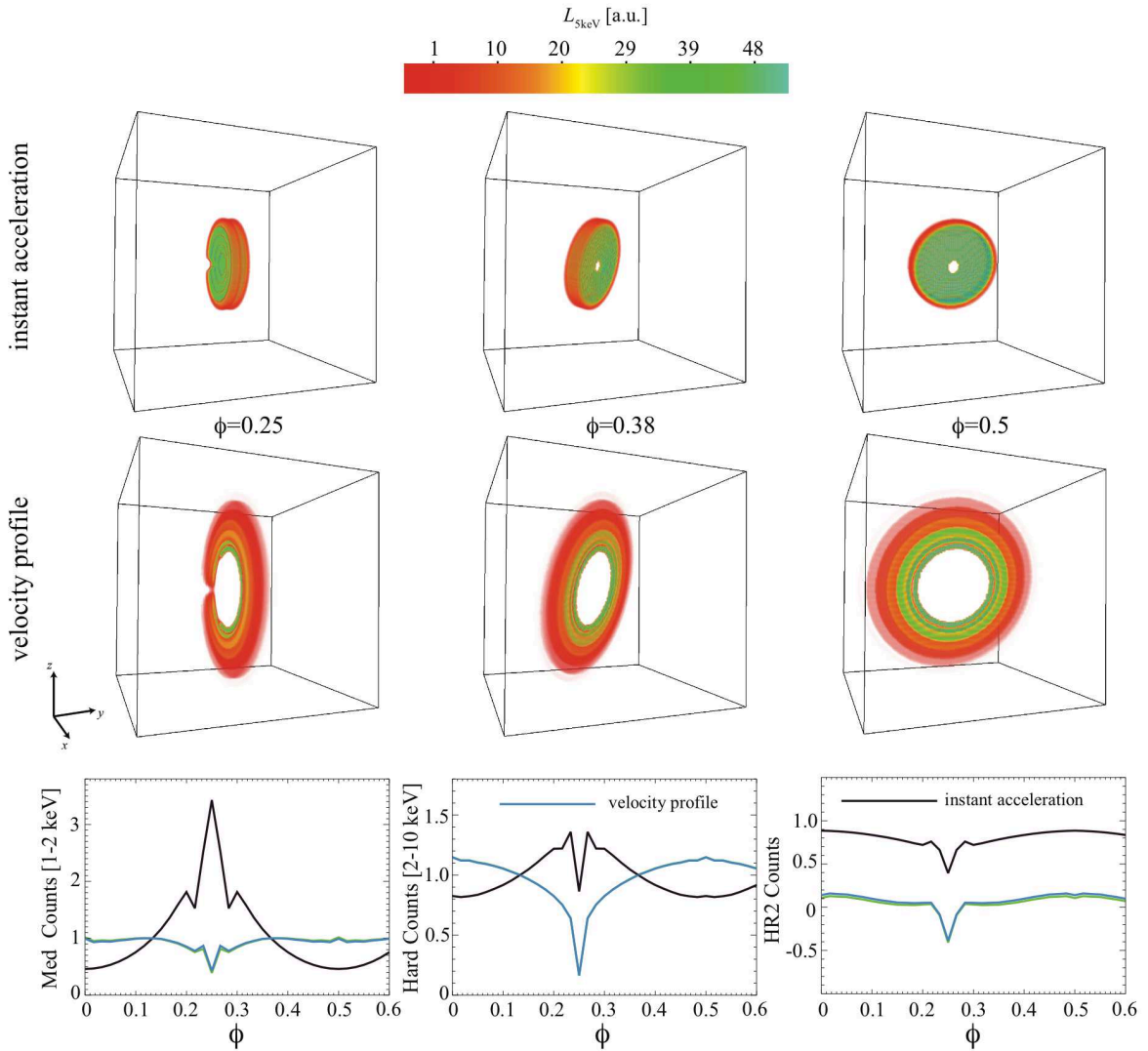}
\caption{Intensity maps (in arbitrary units) at 5~keV for model the {\it instant acceleration} and {\it velocity profile}  models at orbital phases $\phi=$ 0.25, 0.38, and 0.5 (from left to right). {\it Bottom Panels}: Lightcurves at medium ({\it left}) and hard ({\it middle}) bands, and for the hardness ratio  ({\it right}) as predicted from  the {\it instant acceleration} and {\it velocity profile}  models. Counts are 5~ks bins. Lightcurves are normalized to $\phi=0.37$, between eclipsing and conjunction phases.  The {\it green} lines shows the lightcurves for the case in which the opacity  of the unshocked wind material was calculated using  the   stellar blackbodies as  illuminating sources.  To examine how X-ray radiation affects the absorption cross section, the {\it bottom} panels also includes a model ({\it blue} lines) with an added  bremsstrahlung  component  ($T_{\rm ff}= 3 \times 10^7$~K and  $L_{\rm ff} =10^{34}$~erg$\,$s$^{-1}$). Here  $T_{\rm ff}$ and  $L_{\rm ff}$ were  chosen to closely match the total unabsorbed X-ray radiation, although  we found  that the opacity of the wind at the energy ranges where the lightcurves are constructed is not appreciably  altered by the specific spectrum and intensity of the illuminating X-ray source. As can be clearly seen, there is no discernible differences between the {\it green} and {\it blue} lines in the  hard band and only minute differences in the medium band.}
\label{fig:montage}
\end{figure*}

In this way, in the central regions of the shocked wind ($y/x_0 < 4$), the  temperature in the  {\it velocity profile}  model remains  below $\sim 10^6$~K (Figure~\ref{fig:prof}), such that this material does not contribute to the X-ray emission at energies above  0.1~keV.  The main contribution to the X-ray luminosity in the  {\it velocity profile}  model   thus arises  from layers  within the shocked wind  region  located at distances that are significantly  larger  than the binary separation and, as a result, are not eclipsed by the stars.  What is more, 
due to the clear temperature inversion  in the  {\it velocity profile}  model, emission at softer X-ray energies  (which is more readily  absorbed by the stellar wind gas) is produced  in layers that are located  deeper within the shocked wind region that those emitting  higher energy X-rays.  As we will discuss in Section~\ref{sec:light}, both of these effects have  important  consequences on the resulting X-ray lightcurves.

\subsection{X-ray Lightcurves}\label{sec:light}

The behavior of the lightcurves depends on the thermodynamical structure of the shocked  region and the opacity  of the unshocked wind material.
The  lightcurves for the medium ($M$: 1-2~keV) and hard ($H$: 2-10~keV) bands, and for the hardness ratio $HR_2=(H-M)/(H+M)$ are calculated 
by  ray-tracing the  emission  from the shocked wind region over the three-dimensional gas maps. 
In Figure \ref{fig:montage} we compare the intensity of the emitting regions contributing to the observed luminosity at  5~keV for the two different wind initiation models. 
In the case of the {\it velocity profile}  model,  the hot gas contributing with the bulk of the X-ray luminosity  is no longer confined midway between the stars (top row) but it is located in an extended,  ring-like structure (middle row) and, as a result, the emission is not  drastically absorbed by the stars and their close-by  winds.  In the {\it instant acceleration} model, on the other hand, the confinement of the emitting region midway between the stars  causes the X-ray lightcurve to show prominent eclipses ($\phi=0,0.5$). 

In the  {\it velocity profile}  model, the majority of the soft X-ray emission arises in  layers that are deeper within the shocked region  than  in the {\it instant acceleration} model (Figure~\ref{fig:prof}). This  causes  more efficiently  absorption at all orbital phases, which  is reflected in the shape of the resulting  lightcurve, which in the medium X-ray band,  becomes much flatter than in the {\it instant acceleration} model (Figure \ref{fig:montage}). In addition, the density in the shocked region is significantly  increased in the {\it velocity profile}  model,   causing  a sizable increase in the total column density along the contact discontinuity. This increase in optical depth  causes a sharp  decrease in the X-ray luminosity when  $\phi=0.25$ at both  hard and medium energies. This  decrement  is then  followed by a steady  increase in X-ray luminosity as the optical eclipses are taken place ($\phi=0,0.5$) and  is more pronounced at medium X-ray  energies (Figure \ref{fig:montage}).

It is important to note  that in the  {\it instant acceleration} model, because the softer X-ray emission arises from more distant  (and less opaque) layers within the shocked region (Figure~\ref{fig:prof}), the medium X-ray band light curve shows a sizable increase in luminosity at $\phi=0.25$ when softer photons emitted from the shocked wind  region can  reach the observer without transversing the cold,  more opaque wind. As a consequence of both the decrease in  temperature and   increase in  absorption within the shock wind region, the global emission becomes softer  in the {\it velocity profile}  model  as can be seen in the  comparison  of the hardness ratio depicted  in Figure \ref{fig:montage}.  The total observed luminosities in the  1-10~keV energy range  for the {\it instant acceleration} and  {\it velocity profile}  models are $L_X=1.13\times10^{35}(d/8~{\rm kpc})^2\;$erg$\,$s$^{-1}$ and $L_X=1.86\times10^{34}(d/8~{\rm kpc})^2\;$erg$\,$s$^{-1}$, respectively. Having laid the foundation for how the size and structure of the shocked and unshocked regions  are shaped  by the wind  initiation  properties, we now turn our attention to  whether the simulated flow properties can robustly explain salient X-ray observational features of WR~20a. 

\begin{figure}[]
\centering
\centering\includegraphics[height=1.4\linewidth,clip=true]{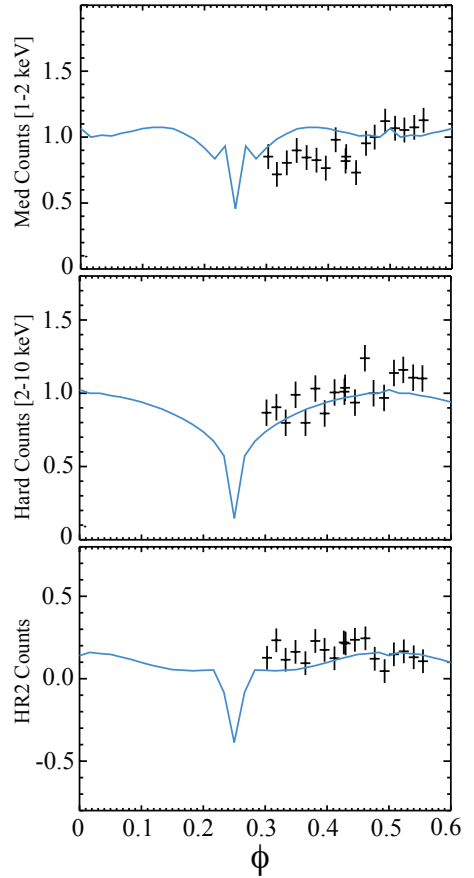}
\caption{Comparison between lightcurve predicted by the {\it velocity profile} model (black line) and the observed X-ray lightcurve reported by \citet{naze08}. Medium, hard and hardness ratio (from top to bottom) are displayed. Counts are 5~ks bins. Lightcurves are normalized to the $\phi=0.47$, the intermediate phase at which observations were taken. The total observed  luminosity in the  1-10~keV energy range predicted for this model is  $L_X=1.8 \times10^{34}(d/8~{\rm kpc})^2\;$erg$\,$s$^{-1}$. This should be compared to $L_{X}= 5.17=\times 10^{33}(d/8~{\rm kpc})^2$ erg$\,$s$^{-1}$, the unabsorbed luminosity  derived by \citet{naze08}, although caution must be taken as this estimate was derived assuming  only Galactic  absorption.}
\label{fig:xray}
\end{figure}

\section{Discussion}
\label{sec:discussion}
In this work, we have presented a simplified  formalism for modeling the relative complexity of the stellar wind  initiation and  interaction  in the massive, close binary WR~20a. Despite its simplicity, this model can successfully  reproduce the observations. In Figure~\ref{fig:xray},  we compare our {\it velocity profile}  model predictions  to the  {\it Chandra} lightcurves  of WR 20a   obtained by \citet{naze08}. Overall, the predicted X-ray lightcurves show no significant flux decrement during secondary optical eclipse but instead appear more luminous and softer as the shock-wind region is observed face on. Notably, such features are only present in models when the emanating wind is assumed to be accelerated (Figure~\ref{fig:montage}).  This system  provides a direct diagnostic of the stellar wind initiation  and suggests that more detailed observations  of WR~20a could help to constrain answers related to  the currently unsolved problem of how WR stars can drive such strong winds \citep{owocki2011}.

As observed in Figure~\ref{fig:xray}, our models not only explain current {\it Chandra} observations but provide detailed predictions over the entire orbital phase.  With a more precise  estimate of $d$ (the distance to the source) and  a better orbital phase monitoring characterization, the approach presented here may be extended to explore several interesting questions. In particular, we have considered the idealized case in which the structure of the shocked wind region is not affected by the  orbital motion, we  have assumed that the wind profile is characterized by $\beta=1$  \citep{rauw05} and have neglected the additional pressure provided by the WR-star light, which is expected to lead to a  radiative braking of the impacting WR wind.  As observations improve, three-dimensional simulations of WR~20a with different velocity profiles  will be needed to test  two-dimensional results and analyze how projection effects can alter our interpretation of both structures and lightcurves.

While it is important to investigate the behavior of individual objects, we should not lose sight of the common physical processes involved. 
There has been detailed hydrodynamical models aimed at understanding the X-ray lightcurves  of  WR+O stars:  $\eta$~Carinae \citep{parkin09,okazaki08,corcoran10}, WR~140 \citep{corcoran11} and WR~22 \citep{parkin11}, which in some case have  separations of only a few O-star radii. Generally the WR wind in these systems is so much stronger that it is expected to overwhelm its companion's outflow. Orbital phase monitoring of such systems suggests that contrary to what a simple hydrodynamic ram balance between the two stars might suggest  the wind-wind interface is generally kept away from the O-star surface, possibly due to the additional pressure provided by the O-star light impacting the WR wind  \citep{gayley97,parkin09}.  It is  of course interesting to study these systems as they appear to be rather common, but it would be better to use simpler systems like WR~20a as proving grounds of the relative complexity of WR wind initiation.

\acknowledgments 
We thank A. Bonanos, M. Krumholz, L. Lopez,  A. MacFadyen, J. Maund, S. Owocki, A. Raga, A. Rosen and A. Vigna-G\'omez  for helpful discussions. We also thank the referee for providing constructive comments and help in improving the contents of this paper. Computations were performed on the UCSC Hyades and Laozi computer clusters. We acknowledge support from the David and Lucille Packard Foundation, NSF grants MRI-1229745 and AST-0847563, DGAPA-PAPIIT-UNAM grant IA101413-2, and CSIC JAE-PREDOC fellowship.


\begin{thebibliography}{}

\bibitem[Abbott et al.(1986)]{abbott86} Abbott, D.~C., Beiging, J.~H., Churchwell, E., \& Torres, A.~V.\ 1986, \apj, 303, 239 

\bibitem[Antokhin et al.(2004)]{antokhin04} Antokhin, I.~I., Owocki, S.~P., \& Brown, J.~C.\ 2004, \apj, 611, 434 

\bibitem[Bonanos et al.(2004)]{bonanos04} Bonanos, A.~Z., Stanek, K.~Z., Udalski, A., et al.\ 2004, \apjl, 611, L33 

\bibitem[Cappa et al.(2004)]{cappa04} Cappa, C., Goss, W.~M., \& van der Hucht, K.~A.\ 2004, \aj, 127, 2885 

\bibitem[Castor et al.(1975)]{castor75} Castor, J.~I., Abbott, D.~C., \& Klein, R.~I.\ 1975, \apj, 195, 157 

\bibitem[Chlebowski et al.(1989)]{chlebowski89} Chlebowski, T., Harnden, F.~R., Jr., \& Sciortino, S.\ 1989, \apj, 341, 427 

\bibitem[Corcoran et al.(2010)]{corcoran10} Corcoran, M.~F., Hamaguchi, K., Pittard, J.~M., et al.\ 2010, \apj, 725, 1528 

\bibitem[Corcoran et al.(2011)]{corcoran11} Corcoran, M.~F., Pollock, A.~M.~T., Hamaguchi, K., \& Russell, C.\ 2011, arXiv:1101.1422 

\bibitem[De Colle \& Raga(2006)]{decolle06} De Colle, F., \& Raga, A.~C.\ 2006, \aap, 449, 1061 

\bibitem[De Colle et al.(2008)]{decolle08} De Colle, F., Raga, A.~C., \& Esquivel, A.\ 2008, \apj, 689, 302 

\bibitem[De Colle et al.(2012)]{decolle12} De Colle, F., Granot, J., L{\'o}pez-C{\'a}mara, D., \& Ramirez-Ruiz, E.\ 2012, \apj, 746, 122 

\bibitem[Dalgarno \& McCray(1972)]{dalgarno72} Dalgarno, A., \& McCray, R.~A.\ 1972, \araa, 10, 375 

\bibitem[Ferland et al.(2013)]{ferland13} Ferland, G.~J., Porter, R.~L., van Hoof, P.~A.~M., et al.\ 2013, RMxAA, 49, 137 

\bibitem[Gr{\"a}fener \& Hamann(2005)]{grafener05} Gr{\"a}fener, G., \& Hamann, W.-R.\ 2005, \aap, 432, 633 

\bibitem[Gayley et al.(1997)]{gayley97} Gayley, K.~G., Owocki, S.~P., \& Cranmer, S.~R.\ 1997, \apj, 475, 786 

\bibitem[Lucy \& White(1980)]{lucy80} Lucy, L.~B., \& White, R.~L.\ 1980, \apj, 241, 300 

\bibitem[Montes et al.(2009)]{montes09} Montes, G., P{\'e}rez-Torres, M.~A., Alberdi, A., \& Gonz{\'a}lez, R.~F.\ 2009, \apj, 705, 899 

\bibitem[Naz{\'e} et al.(2008)]{naze08} Naz{\'e}, Y., Rauw, G., \& Manfroid, J.\ 2008, \aap, 483, 171 

\bibitem[Naz{\'e}(2009)]{naze09} Naz{\'e}, Y.\ 2009, \aap, 506, 1055 

\bibitem[Okazaki et al.(2008)]{okazaki08} Okazaki, A.~T., Owocki, S.~P., Russell, C.~M.~P., \& Corcoran, M.~F.\ 2008, \mnras, 388, L39 

\bibitem[Owocki(2011)]{owocki2011} Owocki, S.\ 2011, Bulletin de 
la Societe Royale des Sciences de Liege, 80, 16 

\bibitem[Pittard \& Corcoran(2002)]{pittard02} Pittard, J.~M., \& Corcoran, M.~F.\ 2002, \aap, 383, 636 

\bibitem[Pittard \& Parkin(2010)]{pittard10} Pittard, J.~M., \& Parkin, E.~R.\ 2010, \mnras, 403, 1657 

\bibitem[Parkin et al.(2009)]{parkin09} Parkin, E.~R., Pittard, J.~M., Hoare, M.~G., Wright, N.~J., \& Drake, J.~J.\ 2009, \mnras, 400, 629 

\bibitem[Parkin \& Gosset(2011)]{parkin11} Parkin, E.~R., \& Gosset, E.\ 2011, \aap, 530, A119 

\bibitem[Pittard(2009)]{pittard09} Pittard, J.~M.\ 2009, \mnras, 396, 1743 

\bibitem[Pollock et al.(2005)]{pollock05} Pollock, A.~M.~T., Corcoran, M.~F., Stevens, I.~R., \& Williams, P.~M.\ 2005, \apj, 629, 482 

\bibitem[Rauw et al.(2005)]{rauw05} Rauw, G., Crowther, P.~A., De Becker, M., et al.\ 2005, \aap, 432, 985 

\bibitem[Sana et al.(2006)]{sana06} Sana, H., Gosset, E., \& Rauw, G.\ 2006, \mnras, 371, 67 

\bibitem[Skinner et al.(2012)]{skinner12} Skinner, S.~L., Zhekov, S.~A., G{\"u}del, M., Schmutz, W., \& Sokal, K.~R.\ 2012, \aj, 143, 116 

\bibitem[Stevens et al.(1992)]{stevens92} Stevens, I.~R., Blondin, J.~M., \& Pollock, A.~M.~T.\ 1992, \apj, 386, 265 

\bibitem[Usov(1992)]{usov92} Usov, V.~V.\ 1992, \apj, 389, 635 

\end{thebibliography}
\end{document}